# FaultNet: A Deep Convolutional Neural Network for bearing fault classification

Rishikesh Magar[1], Lalit Ghule[1], Junhan Li[1,2], Yang Zhao[1,3], and Amir Barati Farimani[1]

[1]Mechanical Engineering Department, Carnegie Mellon University, Pittsburgh, PA 15213 USA
[2]Electrical and Computer Engineering Department, Carnegie Mellon University, Pittsburgh, PA 15213 USA
[3]Civil and Environmental Engineering Department, Carnegie Mellon University, Pittsburgh, PA 15213 USA
Corresponding author: Amir Barati Farimani (e-mail: barati@cmu.edu).

**ABSTRACT** The increased presence of advanced sensors on the production floors has led to the collection of datasets that can provide significant insights into machine health. An important and reliable indicator of machine health, vibration signal data can provide us a greater understanding of different faults occurring in mechanical systems. In this work, we analyze vibration signal data of mechanical systems with bearings by combining different signal processing methods and coupling them with machine learning techniques to classify different types of bearing faults. We also highlight the importance of using different signal processing methods and analyze their effect on accuracy for bearing fault detection. Apart from the traditional machine learning algorithms we also propose a convolutional neural network FaultNet which can effectively determine the type of bearing fault with a high degree of accuracy. The distinguishing factor of this work is the idea of channels proposed to extract more information from the signal, we have stacked the 'Mean' and 'Median' channels to raw signal to extract more useful features to classify the signals with greater accuracy.

## I. INTRODUCTION

With the advent of the 4th industrial revolution, industries across the globe are using artificial intelligence (AI) to improve their processes and increase efficiency to meet the ever-rising customer demands. In this rapidly changing landscape of technology, organizations across the globe, have increased the presence of sensors on the production floor with the motivation of gathering data that can give them valuable insights into their processes[1]. This sensory data contains rich information about the machine and its effective analyses using AI can contribute significantly towards preventive maintenance, quality control, and increased process efficiency[2]. Realizing these obvious benefits of cost-effective tools like AI, organizations across the world are turning towards smarter technologies.

Driven by the keenness of the industry to embrace advanced digital technologies, many researchers are using different signal processing methods and coupling them with machine learning algorithms to address some of the complicated research problems. For example, Thomazella et al. used digital signal processing techniques such as short-time Fourier transform (STFT) and the ratio of power (ROP) to extract features from vibrations signals captured to monitor chatter phenomenon during the grinding process [3]. In another paper on signal processing on vibration data, Zoltan et al., have demonstrated that signal processing techniques such as Discrete Wavelet and Wavelet Packet Transform are effective in extracting features from the frequency domain for fault detection [4]. Their simulated results proved the techniques are even capable of predicting abnormalities exploring long-term tendencies of the detected signals. Signal processing techniques have been performed on acoustic signals as well. Adam et al. have proposed a signal processing technique named MSAF-RATIO-24-MULTIEXPANDED-FILTER-8. This technique is used on acoustic signals captured from electric motors and extracted features are used to classify motor faults [5].

In this work, we will be focusing on the vibration signal data and will be analyzing the different methods of fault detection in bearings using vibration signals. Vibration data has many applications in the areas of structural weakness or looseness, rotating component looseness, and validating the presence of resonance. The optimal monitoring of vibration signals can thus help the analysis of machine performance more effectively, improve efficiency, and more importantly give us insights about machine health. With bearing failure being one of the major contributors to the downtime of industrial machines, it is very important to address this problem with high reliability and reduce the break-down of machines [6][7][8]. As determined by Zhang et al., for rotating machine health monitoring, vibration signal is very important as it contains rich information regarding machine health [9]. Therefore, the analysis of vibration data may help us in the detection and prevention of faults in bearing. In their study, Samanta et al. used time domain statistical features extracted from vibrations signals to classify faults using an artificial neural network[10]. This study was one of the earliest attempts to utilize the capabilities of deep learning for bearing fault detection using vibration signal data. Apart from signal featurization, some researchers have also used wavelet decomposition to extract relevant information from the signal. In a study conducted on wavelet transformation of vibrations signals for fault diagnosis, Sun W et al., use a combination of discrete wavelet transforms and envelope analysis using which they extract the characteristic spectrum of rolling bearing vibration data. Subsequently, a spectrum cross-correlation coefficient is then applied to identify different operating conditions of rolling bearings[11]. Based on this coefficient, different vibration signals are then classified.

Building upon the previous works, we use different signal featurization methods to extract 14 features from the raw vibration signals to classify bearing faults using machine learning and deep learning approaches. In order to comprehensively analyze the signal data, we also implemented wavelet decomposition on the raw signal and couple it with machine learning approaches to evaluate its performance for bearing fault classification. With the motivation of developing a generalized model, we evaluate our machine learning and deep learning approaches on two major publicly available datasets for bearing fault classification.

The first dataset that we analyze has been developed by Case Western Reserve University (CWRU) bearing center[12]. The dataset from the CWRU bearing center will be referred to as the CWRU dataset throughout the paper. The CWRU dataset is one of the important datasets in this research area and has been widely used by researchers to benchmark the performance of their models. In their study, Smith et al., have proposed a benchmark for the CWRU dataset using three different techniques. They have carefully analyzed the different ball faults and compared the signal data amongst the faults. Thus, articulating the difference among signals data when different types of fault occur. However, they do not use the signal featurization techniques that we have employed and don't use any deep learning models[13]. Many researchers recently have used different deep learning models on the CWRU bearing dataset. In their review paper, Zhang et al., have compiled a comprehensive list of different methods used by researchers working in this area. Based on their review of different methods, it is evident that the best performing deep learning models have accuracies in the range of (97%-99%)[14]. Another recent review paper by Neupane et al., also discusses different bearing fault classification datasets, signal feature extraction techniques, and some of the highly accurate deep learning architectures[15]. Based on both the review papers we can conclude that deep learning methods are highly compatible and effective when addressing the bearing fault diagnostics problem.

Most of the deep learning architectures used for bearing fault diagnosis are based on Convolutional Neural Network (CNN). Guo et al. propose a hierarchical adaptive deep convolution network for bearing fault size prediction. In their paper, they convert the signal data into a 32x32 array and use CNNs to accomplish the task. However, their work does not use other information available from signal data like skewness, kurtosis, impulse factor, RMS value[16]. Another work done by Pham et al. proposes a method that converts the signal data into its spectrogram which is then fed to VGG16 for classification[17], [18]. In their paper, they used only four classes and achieved 98.8% accuracy. When compared to their work, we achieved a comparable accuracy on 10 classes with computationally inexpensive architecture. Pan et al., employed 1D CNN and LSTM, in order to take advantage of the signal data, in their paper one-dimensional CNN and LSTM are combined into one unified structure by using the CNNs output as input to the LSTM to identify the bearing fault types[19]. They also compare the usage of nine different featurization techniques and using them with different traditional machine learning algorithms. However, Pan et al., do not use stacked median and mean channels in their work and use a more computationally heavy framework by combining the CNN and LSTM approaches. Most of the state-of-the-art works report an accuracy of more than 98% in bearing fault detection. Guo et al., in their paper, used Stacked denoising Autoencoders have obtained an accuracy of 99.83%. However, they separate the data as per the size of the fault and then make predictions and have only 4 classes in their predictions[20]. As the dataset created by the CWRU bearing center is able to mimic the actual operating conditions the dataset contains some noisy signals as is expected in the actual environment. Therefore, the use of SDAE has been made particularly by researchers to make their predictions more resilient to the noise in the dataset[21][22]. Another approach used by Li et al. combines the convolutional neural network and Dempster-Shafer theory-based evidence fusion. In their work, they demonstrate adaptability to different loads and report an accuracy of 98.92% [23]. LiftingNet by Pan et al. proposes split, predict and, update blocks that are accurately able to predict the bearing faults and are adaptable to different motor speeds and loads. However, their approach is not able to gauge the size of the fault[24]. Our FaultNet can not only predict the type of fault but also the size of the fault based on the input signal. Wang et al. propose the creation of a time-frequency image of the signal and classifying them with AlexNet based architecture[25][26]. When compared to AlexNet which has 5 convolutional layers ours is a relatively inexpensive architecture computationally. Roy et al. propose an autocorrelation-based methodology for feature extraction from a raw signal and then use the random forest classifier for fault classification. They achieve comparable accuracies to the deep learning methods discussed earlier[27].

The second dataset that we have considered in this work is the Paderborn University Data Center bearing dataset[28]. From here onwards, the dataset from Paderborn University will be referred to as the Paderborn dataset throughout the paper. The dataset has vibration as well as motor current signal captured on the test-rig. In the paper proposing the dataset, to extract the features, Fast Fourier Transform (FFT) and power spectral density (PSD) are performed on vibration and motor current signal. After feature extraction and feature selection,18 features emerge for motor current signals, and 15 features are extracted for the vibration signal data[29]. Using conventional machine learning approaches, Karatzinis et al, achieved the highest accuracy of 98%. However, they do not use advanced deep learning techniques which may possibly increase the accuracy. In another study Zhong et al., transform the signal using Short-time Fourier Transform (STFT) and use CNN to classify the bearing fault [30]. On the transformed signal domain, they apply CNN. The average accuracy achieved is 97.4%. Compared to their work, our model yields better results by directly using the raw signal. Bin Li et al., have implemented 1 dimensional CNN architecture and the best result achieved by them is 98.3 % accuracy in fault classification. However, they have not explored the 2D CNN method to improve the results. In another study, Pandhare et al., have implemented 2D CNN for the bearing fault classification on the Paderborn dataset[31]. In their work, they have demonstrated 2D CNN on 3 different signal types – raw time domain signal, envelope spectrum, and spectrogram. The maximum accuracy achieved is with a spectrogram. For raw signals, the achieved accuracy of 95% is slightly lower when compared to the other studies. Another group of researchers, Wang et al, have proposed a method to use 1D CNN as well as 2D CNN together to predict the fault class in the Paderborn dataset [32]. They have concatenated the 1D CNN output with 2D CNN output before passing it on to a fully connected neural network for classification. Their resultant accuracy for the classification task is 98.58%. However, their approach is computationally expensive and hence may not be very suitable for online deployment.

In this paper, we propose FaultNet, a CNN based model to determine different types of bearing faults with high accuracy. The aim of this paper is to set a benchmark for bearing fault detection using conventional machine learning algorithms and deep learning techniques on CWRU and Paderborn datasets. It is important to note that the base architecture for both the datasets is the same and the performance of FaultNet is not dataset specific, suggesting wide applicability and deployability of the model to detect different types of bearing faults. We achieve state-of-the-art accuracies for both datasets while proposing a different methodology to extract features from the data. We also study different signal processing techniques and compare accuracies of the traditional machine learning algorithms when combining different types of signal features and our own 2D CNN model.

## II. DATASET PREPROCESSING

### A. Case Western Reserve University Bearing Dataset

The test rig to generate the dataset consists of a 2 hp electric motor to the left, driving a shaft on which a torque transducer and encoder are mounted in the middle coupled to a dynamometer in the right. The torque is applied to the shaft via a dynamometer and electronic control system. The test rig also includes bearings at both the drive end (DE) and fan end (FE) of the motor. The bearing at the DE and FE are 6205-2RS JEM and 6203-2RS JEM, respectively. The 6205-bearing used for data collection is a Single Row Deep Groove Radial Ball Bearing with an inner diameter of 25mm, an outer diameter of 52mm, and 15mm in width. To collect the vibration signal data single point faults were artificially induced using electro-discharge machining (EDM) with fault diameters from 7 to 28 mils (0.18 to 0.71mm). The motor loads varied from 0 to 3 hp (approximate motor speeds of 1720 to 1797 rpm). The vibration data was collected using accelerometers, which were attached to the housing with magnetic bases. The data was collected with two sampling frequencies, one with 12,000 samples per second, and 48,000 samples per second, and was processed using MATLAB®. In their study, the DE & FE bearing data for the normal (N), inner race fault (IF), outer race fault (OF), and the rolling element(ball) fault (BF) conditions was acquired for fault pattern classification where the fault diameters were selected to be 7 mils, 14 mils, and 21 mils.

TABLE I: CWRU BEARING HEALTH CONDITIONS AND CLASS LABELS

| Health Condition | Fault size (mm) | Total dataset | class labels |
|---|---|---|---|
| Normal | - | 280 | 0 |
| ball fault | 0.18 | 280 | 1 |
| ball fault | 0.36 | 280 | 2 |
| ball fault | 0.53 | 280 | 3 |
| inner race fault | 0.18 | 280 | 4 |
| inner race fault | 0.36 | 280 | 5 |
| inner race fault | 0.53 | 280 | 6 |
| outer race fault | 0.18 | 280 | 7 |
| outer race fault | 0.36 | 280 | 8 |
| outer race fault | 0.53 | 280 | 9 |

Ten different conditions are investigated to verify the accuracy of the proposed method in consideration of multiple fault patterns. The vibration signals of ten health conditions are referred to in table 1. In this paper, we used the data from the drive end of the test rig. The sampling frequency chosen is 48 kHz with the load condition being 2 HP at 1750 rpm. To analyze and classify different bearing faults we do some preprocessing steps on the dataset. The rotating speed of the shaft is 1750 rpm and the sampling frequency is 48 kHz implies that approximately 1670 data points will be collected for one revolution. Out of 1670 data points, the first 35 points and last 35 points are ignored to account for the noise in the data. Thus, 467600 data points of each fault class are chosen and divided into 280 samples, with 1670 data points. Finally, we have 2800 samples with 10 different classes with 280 samples each. Further details, which introduce the test set-up and other data collected, can be found at the CWRU Bearing Data Center website.

### B. Paderborn University Dataset

This dataset is generated using 32 bearings. The bearing type used for this dataset generation is 6203, which stands for Deep Groove Ball Bearings with dimensions (inner diameter, outer diameter, and width) – 17X40X12mm. Out of 32, 6 bearings are healthy, 12 bearings have artificially created defects and the remaining 14 bearings are naturally damaged. The artificial defects have been created by using drilling, EDM, and electric engraving machine. The artificial defects are produced on both, inner and outer race. The natural damages are produced by accelerated lifetime tests. A detailed description can be found in the paper[28].

Further, the bearings' samples can be divided into 3 classes, healthy, inner race fault, and outer race fault. By this classification, there are 6 healthy bearings, 11 inner race fault bearings, and 12 outer race fault bearings. This amounts to 29 bearings in total. The Remaining 3 bearings are omitted due to their nature of the fault. These 3 bearings have inner as well as outer race fault. In the study conducted by Paderborn University, the authors have classified these bearings on the grounds of the maximum contributing fault. If the inner race damage is more compared to the outer race, the bearing is classified as inner race fault bearing. For the current model, we used 29 bearings data which can be classified distinctively. The data set is generated with multiple combinations of rpm, torque, and load. For the purpose of this study, we use the following combination. N=1500 rpm, load torque=0.7 Nm and Radial force=1000 N.

Each bearing is used 20 times to generate 20 signals with one fixed combination. The signal generated is a vibration signal for 4 sec with a sampling frequency of 64kHz. That means, in a signal, there are 256,000 data points. To avoid initial and ending noise and disturbance, the sample signal is clipped off for the first 1/16th part and the last 1/16th part. Eventually, the signal used has 2,24,000 data points which are used further for featurization. In total 2320 signals have been used for classification.

## III. FEATURIZATION

In every machine learning process, feature engineering plays a very important role and can significantly affect the performance of an algorithm. Feature engineering can directly help the machine learning algorithm to identify the underlying patterns and effectively improve the accuracy of the model. For signal data, featurization includes deriving different domains' features from raw signals such as time domain, time-frequency domain, etc. The vibration signals from machinery components are in general considered to be non-stationary. The non-stationary signals mean that the frequencies present in a signal vary with time [33]. Therefore, it is important to extract features from the time domain as well as the time-frequency domain to capture the time-varying nature of frequencies present in a signal. In this paper, the features extracted from raw signal data include multiple time and time-frequency domain features. Some of the statistical time domain features that we extract include mean, variance, standard deviation, root mean square (RMS). Moreover, features such as kurtosis and skewness are also extracted as these signals are not stationary. In their paper, Caesarendra et al., give us some physical insights into the features as they report the approximate values of kurtosis and skewness for a normal bearing to be 3 and 1 respectively [34]. Hence, for bearings that are faulty, we expect to have kurtosis and skewness values shifted from 3 and 1. Another important observation we made was that for faulty bearings, the bearing signal amplitude undergoes abrupt changes when rolling elements pass over the defective region of the bearing. These abrupt changes are responsible for disturbing the overall distribution of signal and therefore can act as an important clue in detecting faulty bearings. Generally, the value of kurtosis increases and skewness may change to the negative or positive side for faulty bearings. Apart from these features, dimensionless features such as crest factor, shape factor, impulse factor are also extracted. The shape factor is affected by the shape but is independent of the dimension. The crest factor is a measure of an impact when a rolling element comes in contact with the raceway. Table 2 summarizes all the 14 features extracted from the raw signal data along with their mathematical formulae used.

TABLE II: FEATURES AND THE MATHEMATICAL FORMULAE USED TO CALCULATE VALUES FOR EACH SIGNAL

| No | Feature | Formula |
|---|---|---|
| 1 | Mean | $Mean = \frac{1}{n} \sum_{i=1}^{n} x_i$ |
| 2 | Absolute mean | $Abs\ Mean = \frac{1}{n} \sum_{i=1}^{n} |x_i|$ |
| 3 | Maximum | $Maximum = max(x_i)$ |
| 4 | Minimum | $Minimum = min(x_i)$ |
| 5 | Peak to Peak | Maximum - Minimum |
| 6 | Absolute max | $Abs\ Max = max(|x_i|)$ |
| 7 | Root Mean Square | $RMS = \sqrt{\frac{1}{n} \sum_{i=1}^{n} x_i^2}$ |
| 8 | Variance | $Var = \frac{1}{n} \sum_{i=1}^{n} (x_i - \bar{x})^2$ |
| 9 | Clearance factor | $ClF = \frac{Absolute\ max}{\left(\frac{1}{N} \sum_{i=1}^{N} \sqrt{|x_i|}\right)^2}$ |
| 10 | Kurtosis | $Kurt = \frac{\sum_{i=1}^{n} (x_i - \bar{x})^4}{n \times var^2} - 3$ |
| 11 | Skewness | $= \frac{\frac{1}{n} \sum_{i=1}^{n} (x_i - \bar{x})^3}{\left(\sqrt{\frac{1}{n} \sum_{i=1}^{n} (x_i - \bar{x})^2}\right)^3}$ |
| 12 | Impulse Factor | $IF = \frac{Abs\ Max}{\frac{1}{n} \sum_{i=1}^{n} |x_i|}$ |
| 13 | Crest Factor | $CF = \frac{Abs\ Max}{RMS}$ |
| 14 | Shape Factor | $SF = \frac{RMS}{\frac{1}{n} \sum_{i=1}^{n} |x_i|}$ |

Time-frequency domain representation methods such as short-time Fourier transform (STFT), wavelet transform, and Wigner-Ville distribution (WVD) are commonly used for the non-stationary or transient signal. These methods implement a mapping of one-dimensional time-domain signals to a two-dimensional function of time and frequency. The objective is to provide a true time-frequency representation of a signal. Similar, to the methods presented in the review [35] conducted by Feng et al., on time-frequency analysis methods for machinery fault diagnosis, we decomposed these raw signals by employing wavelet decomposition package (WPD) using Haar wavelet as a mother wavelet for the extraction of time-frequency domain features. The wavelet decomposed signal consists of approximation coefficients and detailed coefficients. In this work, we use the approximation coefficients as they are more sensitive towards bearing conditions as suggested in [28] for the extraction of statistical features mentioned in table II.

## IV. RESULTS WITH SIGNAL FEATURIZATION

### A. Case Western Reserve University Bearing Dataset

We used all the 14 features in Table 2 and evaluated the performance of different shallow learning algorithms. The train-test split used was 80-20% and average 5-fold cross-validation accuracy has been reported. Amongst all the models we tried, the random forest yielded the highest accuracy (figure 2(a)). An important functionality of the random forest algorithm is that it provides feature importance which gives the user important insights about the features. The feature importance score of the top 5 features calculated using Random Forest has been demonstrated in figure 1(a). It was observed that absolute mean, variance, RMS, shape factor, and the mean are the 5 most important signal features. Subsequently, to analyze the effects of these important features on the accuracy multiple combinations of these features were used to classify the bearing faults. As expected, the accuracy improves when a model is trained with a greater number of features (figure 2(a)). We also implemented wavelet decomposition on the raw signal data for the CWRU dataset extracted the same 14 features from Table 2 on the decomposed signal. Random forest was the best performing algorithm, and it was observed that wavelet decomposition level 2 and level 3 had a slightly higher performance (figure 3(a)). After testing multiple algorithms and different signal featurization techniques it was observed that the performance did not improve beyond 90% accuracy, to further improve the accuracy we then tried deep learning approaches.

### B. Paderborn University Dataset

Like the CWRU dataset, we used a train-test split of 80-20% and report the average five-fold cross-validation accuracy. For the Paderborn data, the random forest algorithm gave the highest accuracy. Utilizing the feature importance functionality of random forest, the top five important features were calculated. It is important to note that, out of the top five important features, four features are common in the CWRU dataset (figure 1(b)). Similar to CWRU multiple combinations of different features were tried and the accuracy of different shallow learning methods was evaluated (figure 2(b)). We also tested for three different decomposition levels to check the effect of wavelet decomposition on the overall accuracy of the model. As demonstrated in figure 3(b), it is observed that there is a slight increase in accuracy with the decomposition level. Similar to the CWRU dataset we decided to use deep learning to improve the accuracy further.

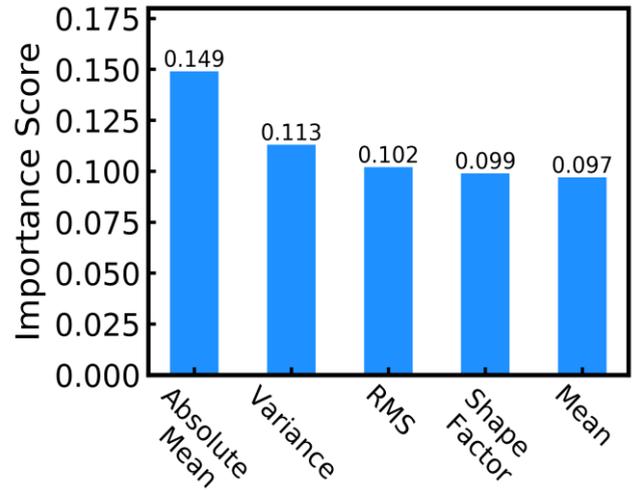

(a)

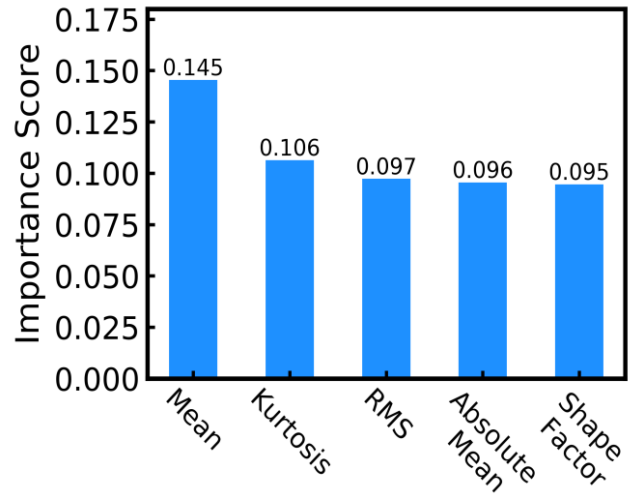

(b)

**FIGURE 1.** Feature importance based on Random Forest results. (a) shows top 5 important features obtained on CWRU bearing dataset. Similarly (b) shows the results obtained for Paderborn University dataset. For both the datasets, out of 5 important features.

After analyzing the results from conventional machine learning approaches with signal featurization, we realized that the best accuracy achieved was not comparable with the state-of-the-art results discussed earlier. Therefore, we decided to build a Convolutional Neural Network (CNN) that is computationally inexpensive and also achieves higher accuracy.

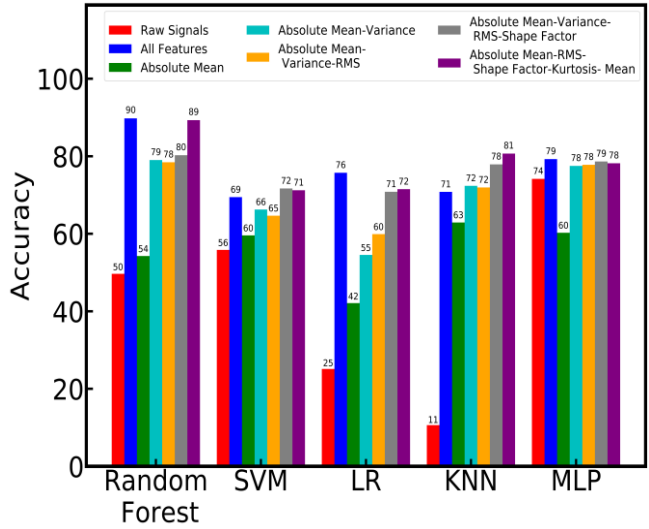

(a)

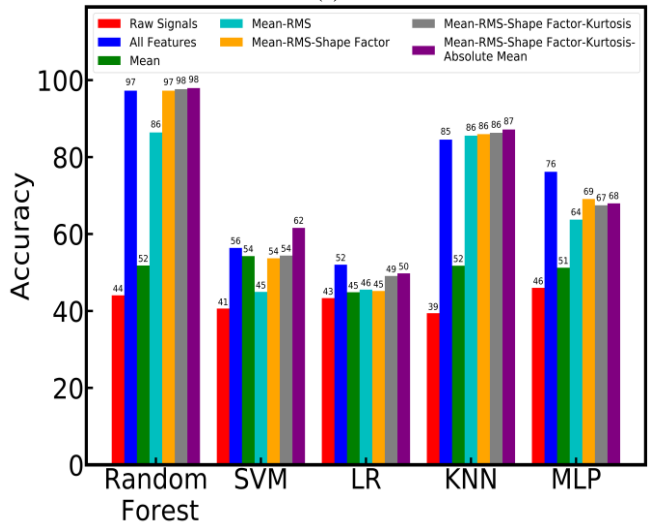

(b)

**FIGURE 2.** Classification accuracy with raw data, all features, and five important features (a) shows accuracy for CWRU dataset for five different classification algorithms. It is evident from the figure that accuracy improves as number of features increases. Only using raw data yields the lowest accuracy.
Similarly (b) shows the results obtained for Paderborn University dataset.

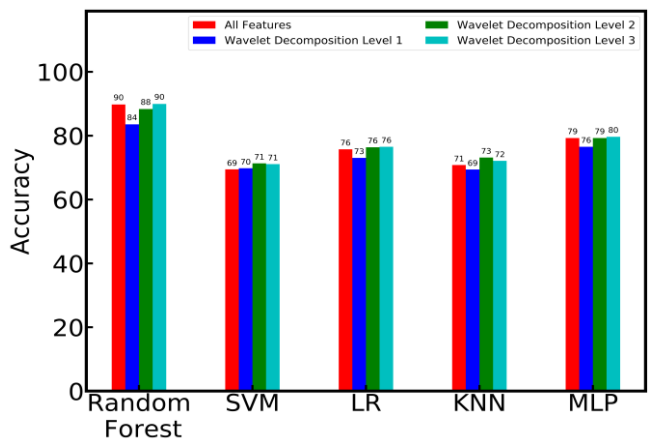

(a)

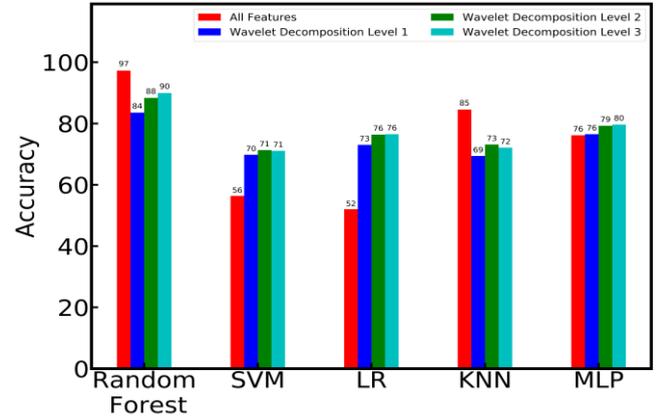

(b)

**FIGURE 3.** Classification accuracy using all 14 features on different signal wavelet decomposition level (a) shows accuracy for CWRU dataset. Decomposing signals at different result in slightly higher accuracy. Also, for some ML algorithms decomposed signals gives better accuracy compared to raw signals.
Similarly (b) shows the results obtained for Paderborn University dataset.

## V. CONVOLUTIONAL NEURAL NETWORK

To further improve the classification performance, we developed FaultNet a CNN based architecture that takes raw signal data as input without any pre-processing. CNNs, because they possess a special ability to extract relevant features from the data, given a task of prediction. In the FaultNet architecture, there are two parts, the first part being the convolution part and the second part is fully connected layers. We have two convolution layers. Each convolution layer is followed by a max-pooling layer. The activation function used for both convolution layers is 'ReLU'. Usage of max-pooling layers ensures that the most important features are selected. The addition of pooling has also led to decreased computational times making the FaultNet relatively inexpensive architecture.

As we know, the raw signals CWRU dataset contains 2800 signals of 1600 data points. The signal data is converted into a 2D array of shape $40 \times 40$. Therefore, we have 2800 signals in the form of 2D arrays of shape $40 \times 40$. Similarly, for the Paderborn University dataset, signals of 250,000 datapoints are split into 100 smaller signals of shape $50 \times 50$ 2D arrays. The convolution operation is performed on the 2D data. The convolution operation is defined by:

$$y(t) = x(t) * w(t)$$

After the convolution layers, the output is flattened and fed to fully connected layers. The fully connected layer has an input layer with 5184 neurons for the Paderborn dataset. There is only one hidden layer with 256 units. According to the dataset, the output layer has either 3 (Paderborn dataset) or (CWRU dataset) 10 neurons. For fully connected layers as well, the activation function used is 'ReLU'. Besides, for the final layer, a drop-out of 0.25 is added to prevent the overfitting in the neural network. Soft-max activation is applied to the outputs of the neurons in the final layer[36] [37]. The overall network architecture is shown in figure 4.

As the problem's nature is classification, we use Cross-entropy loss. For training, we use the 'Adam' optimizer with a constant learning rate of 0.001. The whole network is trained for 100 epochs with a batch size of 128 on NVIDIA RTX2080 GPU. At the end of the training, the loss value for the Paderborn dataset is 0.0003.

## VI. RESULTS WITH CONVOLUTIONAL NEURAL NETWORK

Initially, we tried to predict the bearing fault class with only raw signals. However, the 5-fold accuracy maxed out at 95.27%. With the state-of-the-art model achieving accuracies in the range of 97-100%. We devised a methodology to incorporate more signal information so that the model can learn the signal features better. Thus, we came with the idea of using mean and median channels to augment the raw signal in a bid to improve the accuracy. In order to generate new channels, a sliding window with a length of 10 was used as a filter. For every given sample signal data, the filter scans through the whole sample data from the front to the end. To get the same quantity of data points as the sample data, nine '0' were replenished at the end of the sampled signal data and the filter has been set to shift by length 1 for each time. Consequently, a new channel, which had the same shape size as the original channel of sample signal data, is generated by all outputs while the filter completed the data traversal. For the first additional channel, a mean filter was applied to create a mean channel. For the second channel, the mean filter is substituted with a median filter to generate the median channel. We combined new channels with the original channel as the new input for the 2D convolution model. With an increasing number of channels, the accuracy improves simultaneously. The average accuracy has already improved to 98.50% as evident from figure 5(a). Similar improvements in performance were seen on the Paderborn dataset in figure 5(b).

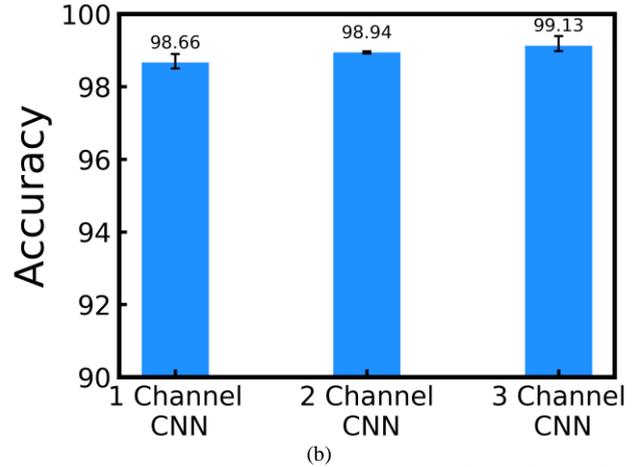

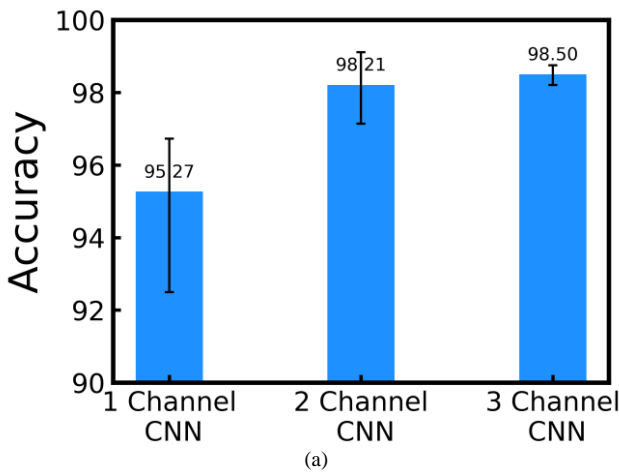

**FIGURE 5.** Comparison of accuracies obtained for different CNN models.
(a) CWRU dataset with one-channel, two channel and three channel approach.
(b) Paderborn dataset with one-channel, two-channel and three channel approach

Apart from the improved accuracy, we observed that the deviation of accuracy was lowered over the five folds of test datasets. Leading us to conclude that channel addition not only improves accuracy but also generates a more stable model in terms of accuracy. In order to augment accuracy as an evaluation metric, we also calculate the precision, recall, and F1 score. The high values of the F1 score for both datasets (Table 3) indicate that the FaultNet architecture is robust and captures the faults with high precision. When comparing the performance of FaultNet with CNN architecture proposed by Zhang et al, we observe that the precision, recall, and F1 score for their method are 0.8155, 0.8105, and 0.8129 [38]. However, these values are only reported with 90 training data points. When we trained the FaultNet with the same number of datapoints we get the precision, recall, and F1 score as 0.799, 0.7924, and 0.7956 respectively. The performance of FaultNet is comparable to the architecture proposed by Zhang et al., even in a low data setting considering the fact that they propose a deeper CNN that has five convolutional layers, and FaultNet has two convolutional layers. Pham et al., have also calculated the precision and recall score on different motor settings and their average scores are 0.9826 and 0.991 respectively [17]. When compared to their architecture FaultNet has a slightly higher precision by 0.0034 and slightly lower recall by 0.0053. The performance of our architecture FaultNet is comparable to other state-of-the-art methods in all performance metrics. Despite being a lightweight CNN architecture, FaultNet is able to achieve comparable results on all performance metrics because of the novelty in the way in which we make use of median and mean channels.

Apart from that, we also plot the confusion matrix for both datasets (Figure 6). We observe that for the CWRU dataset, the model fails to classify datapoints belonging to ball fault with 0.18 mm fault size. The model is confused between ball fault with 0.18 mm fault size and outer race fault with 0.36 mm. This leads us to conclude that the model has a slightly lower accuracy for outer race faults. Interestingly, the model is able to classify smaller size faults with high accuracy. It is evident that class 0 has better accuracy compared to other classes. Class 2 has the highest number of misclassifications. The model has difficulty in distinguishing between the inner race and outer race faults. In general, it is difficult to classify outer race faults as observed from confusion matrixes of both datasets.

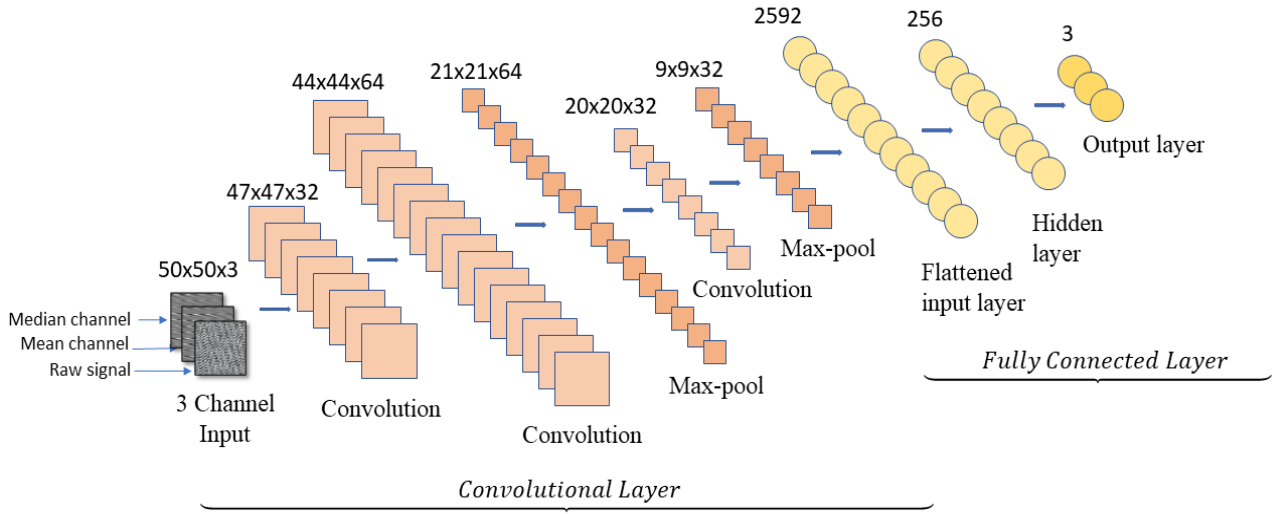

**FIGURE 4.** Convolutional Neural Network Architecture. 2D CNN architecture used for Paderborn University dataset. Different colors represent different operations. There are 2 convolution and 2 max-pool layers along with fully connected neural network with 3 outputs. After each convolution layer, a 'Batch-norm' and 'ReLU' activation is applied.

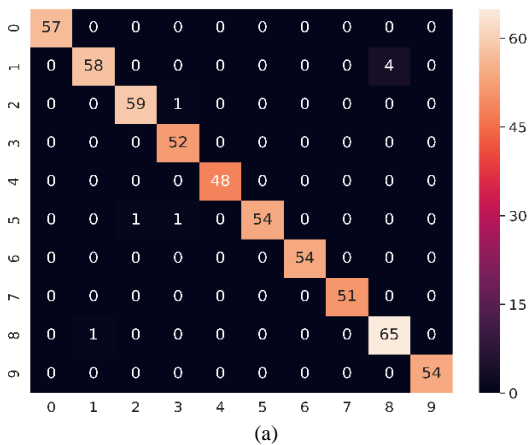

(a)

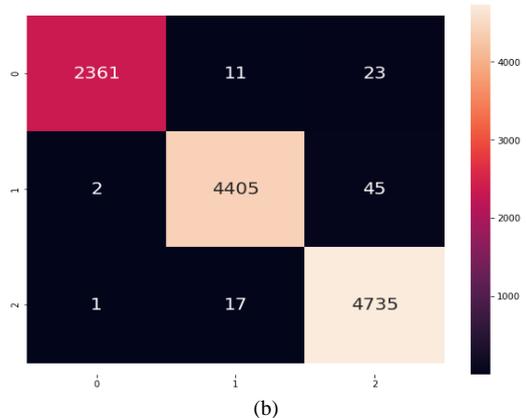

(b)

**FIGURE 6.** Confusion matrix for comparing the inter-class performance.
(a) CWRU dataset: Classes from 0 to 9 correspond to the labels given in table I. The test accuracy corresponding to this confusion matrix is 98.57%.
(b) Paderborn dataset: Class 0 corresponds to healthy bearing. Class 1 and class 2 represent inner race fault bearing and outer race fault bearing, respectively. Overall accuracy is 99.14%.

TABLE III: PERFORMANCE METRICS FOR CWRU AND PADERBORN DATASET

| Metric | CWRU | Paderborn |
|---|---|---|
| Precision | 0.9860 | 0.9906 |
| Recall | 0.9857 | 0.9918 |
| F1 score | 0.9857 | 0.9915 |

## VII. PERFORMANCE EVALUATION FOR NOISE ROBUSTNESS

To evaluate the robustness of the FaultNet architecture and different conventional machine learning algorithms to noise we added white Gaussian noise to the vibration data and assessed the fault classification performance. We chose seven different signal to noise ratios (SNR) to understand how noise affects the performance of the different algorithms. The results for the CWRU and Paderborn datasets are demonstrated in Table IV and Table V, respectively. Among the conventional machine algorithms, we observed that random forest had the highest accuracy and was more robust to noise for both the datasets (CWRU and Paderborn). FaultNet achieves a high accuracy of 97.77% (CWRU) and 98.8% (Paderborn) when the SNR is 10. It can be observed that the accuracy is slightly lower when compared to the original vibration signal. For all the conventional machine learning algorithms and CNN-based FaultNet, accuracy increases as the SNR goes up. It must be noted that for the noisiest signal with SNR equal to –4,

FaultNet performs reasonably well with an accuracy of 82.12% and 89.3% for CWRU and Paderborn respectively. When compared with deep learning-based architecture by Zhang et al., FaultNet achieves comparable accuracy within 1% for SNR values of 8 and 10 and considerably outperforms it when the SNR values are less than 2 on the CWRU dataset [38]. We would like to note that CNN proposed by Zhang et al., consists of 5 convolutional layers whereas our lightweight architecture FaultNet only has 2 convolutional layers, making it more suitable for an online industrial setting. FaultNet is able to achieve high accuracy because of the novel way in which it is able to use information from signals through mean and median channels.

TABLE IV: PERFORMANCE EVALUATION OF FAULTNET FOR NOISE ROBUSTNESS ON CWRU DATASET

| SNR | RF | SVC | LR | kNN | MLP | FaultNet |
|---|---|---|---|---|---|---|
| -4 | 75.16 | 67.14 | 68.75 | 66.25 | 73.64 | 82.12 |
| -2 | 77.27 | 69.01 | 70.14 | 68.32 | 74.89 | 84.14 |
| 0 | 79.04 | 71.52 | 71.71 | 69.39 | 76.03 | 87.68 |
| 2 | 81.65 | 73.24 | 72.17 | 68.75 | 76.21 | 90.74 |
| 6 | 82.38 | 73.89 | 72.78 | 69.25 | 76.42 | 93.17 |
| 8 | 84.29 | 74 | 74.07 | 69.10 | 76.62 | 96.21 |
| 10 | 85.62 | 74.78 | 74.32 | 69.32 | 77 | 97.77 |

TABLE V: PERFORMANCE EVALUATION OF FAULTNET FOR NOISE ROBUSTNESS ON PADERBORN DATASET

| SNR | RF | SVC | LR | kNN | MLP | FaultNet |
|---|---|---|---|---|---|---|
| -4 | 87.73 | 80.72 | 64.42 | 80.63 | 76.12 | 89.3 |
| -2 | 89.27 | 82.47 | 67.97 | 81.96 | 78.41 | 93.3 |
| 0 | 90.04 | 83.86 | 68.68 | 82.62 | 79.03 | 94.1 |
| 2 | 90.65 | 85 | 69.6 | 83.34 | 82.03 | 96.5 |
| 6 | 92.38 | 86.64 | 70.2 | 84.12 | 82.82 | 97.7 |
| 8 | 94.29 | 87.44 | 72.07 | 84.96 | 83.5 | 98.2 |
| 10 | 95.22 | 88.08 | 71.67 | 85.57 | 84.16 | 98.8 |

## VIII. CONCLUSION

In this paper, a systematic approach towards a data-driven vibration-based diagnosis of faults in rolling element bearings is demonstrated. We have benchmarked the performance of different machine learning algorithms by using the featurized signal data and deep learning approaches for the CWRU and Paderborn datasets. Five-fold accuracies of ~99% are obtained for both the datasets indicating the state-of-the-art performance is achieved by the FaultNet architecture. For classification, it is important to have enough differentiating features between classes. As we stack a greater number of channels, the algorithm is able to extract more features compared to single-channel input. Each additional channel could be considered as a feature map of the input which provides more information about the input. This is analogous to grayscale and RGB images. If we convert an RGB image to a grayscale image, often, there happens to be information loss leading to poor performance [39]. Hence, adding more information improves the performance of our model by working exactly opposite to the image conversion from RGB to grayscale.

The novelty of this work is in the concise CNN structure, also, in the concept of augmenting 2D raw signal with its mean and median value channels to extract more meaningful features for CNN. We have demonstrated that the CNN structure devised here improves upon previous methods and has a highly competitive performance compared with state-of-the-art methods. We believe this work can pave the way for online fault detection in the case of bearings which could be extremely beneficial for industries. Our approach can be extended to similar types of datasets.